\documentclass{XrU2005}

\usepackage{graphicx}
\title{RADIO AND X-RAY EMISSION ASSOCIATED WITH THE SUPERNOVA REMNANT 
G352.7-0.1}
\author[1]{E. Giacani}
\author[2]{N. Loiseau}
\author[1]{G. Dubner}
\author[2]{M.J.S. Smith}
\affil[1]{Institute of Astronomy and Space Physics, Buenos Aires, Argentina}
\affil[2]{XMM-Newton Science Operation Center, ESAC, Villafranca del Castillo , Spain}

\begin{document}

\maketitle
\begin{abstract}
We report on new VLA radio and XMM-Newton X-ray observations of the SNR G352.7-0.1.
These high sensitivity, high resolution data reveal that G352.7-0.1 belongs
 to the thermal composite morphological class.
  Small scale structures in radio and X-ray emission are not always correlated and are
  different for the different X-rays bands examined.
The distance to G352.7-0.1 can be constrained between 6.6 and 8.4 kpc.
The study of the HI suggests that G352.7-0.1 is located within a cavity probably
 created by the stellar wind of the precursor star. 
\end{abstract}
\section{Introduction}
G352.7-0.1 is a supernova remnant  (SNR) classified as a shell-like type, 
with a size of $8^{\prime}\times 6^{\prime}$.
 The high resolution VLA image at 1.4 GHz obtained by \citet{Dubner93}
 shows the presence of two overlapping ring structures and an unresolved 
 bright spot over the eastern limb, whose  origin is unclear.
\citet{Kinugasa98} presented an ASCA X-ray image that is
 described as  a shell 
which roughly coincides with the inner radio shell.
 They proposed 
that G352.7-0.1 is a middle-age (2200 years) SNR located at 8.5~kpc,
 evolving in a pre-existing cavity.

We present the comparison between high spatial resolution 
radio data and XMM-Newton X-ray data of G352.7-0.1, carried out  to 
investigate the origin of the observed  morphologies, analysing also their X-ray
spectra. 
 In addition, based on HI data taken from the SGPS 
survey \citep{Mc01} and CO data from the CfA CO 
survey \citep{Dame01} we investigate the atomic and molecular gas 
in the direction to G352.7-0.1 to set constraints on its distance.

\section{Radio and X-ray observations }
The radio image at 4.8~GHz was produced from VLA archival data corresponding
to observations carried out in its DnC configuration.
 The data were processed under the
 Miriad software package following standard procedures. The angular
resolution of the final image is $12^{\prime\prime}\times 9^{\prime\prime}$ and 
the noise is 0.2~mJy/beam.

X-ray images and spectra were obtained from EPIC data of an  
XMM-Newton observation of G352.7-0.1 performed on October 3, 2002.
The public data were extracted from the XMM-Newton Science Archive 
and processed using version 6.5.0 of the XMM-Newton Science Analysis System.
The MOS and pn cameras were operated in FULL FRAME mode with the Medium filter.
Net exposure times were 25~ks and 20~ks for the MOS and pn cameras, respectively. 
 The astrometry of the resulting  
images was confirmed to be accurate to  about 5~arcsec.

\section{Results}
Fig.1 ({\it Left}) shows G352.7-0.1 at 4.8~GHz. The overall appearance of 
 this image  resembles 
 that at 1.4~GHz, but the higher angular resolution of these data reveal clumpy 
structures on small scales. The largest angular scale structures, however, have not been 
fully recovered  because of the incomplete sampling in the 
visibility plane. 

\begin{figure}
      \centering
\vspace{1cm}
      \includegraphics [width=8cm]{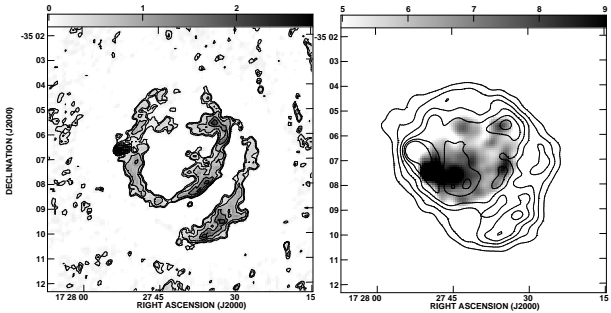}
      \caption{Left: Gray-scale and contour image of G352.7-0.1 at 4.8 GHz.
Right: XMM-Newton X-ray data in the range 0.15 to 8.0 keV (gray-scale) overlaid
with the VLA radio contours at 1.4 GHz (from Dubner et al. 1993).}
\end{figure}

Fig. 1 ({\it Right}) shows the broadband (0.15 - 8.0 keV) EPIC image 
of G352.7-0.1 overlaid with the VLA radio contours at 1.4~GHz as taken
from Dubner et al. (1993). The X-ray emission  is confined within the 
inner radio shell, filling it almost completely. The emission is characterized
by the presence of several knots of emission.
 The
brightest radio spot has not counterpart in the X-ray range and vice-versa. These
new observations reveal that G352.7-0.1 belongs to the composite morphological
class, i.e. shell-like in radio, filled- centre in the X-ray regime.
Narrowband images
centered at the  Si, S, Ar and Fe emission lines are displayed in Fig. 2 
(a), (b), (c) and (d) respectively. The distribution of the emission in 
the energy bands centered at Si and S lines are quite similar to the
broadband image. 
 The image centered in the Ar line, instead, outstands only near the  
 southeastern  X-ray maximum.
 The distribution of the Fe  emission line band 
  is more clumpy and  anti-correlated 
with the other bands.

\begin{figure}
      \centering
\vspace{1cm}
      \includegraphics [width=8cm]{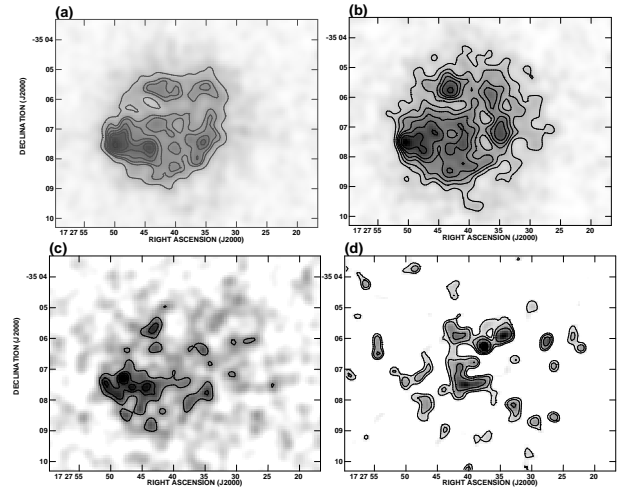}
      \caption{XMM-Newton EPIC images in the bands centered in: (a) the 
      Si XIII/XV line (1.7-2 keV), (b) the S XIII/XV line (2.3-2.6 keV), 
(c) the Ar XVII line (3.1-3.3 keV) and
(d) the Fe K$\alpha$ line (6.3-6.7 keV).}
\end{figure}

EPIC X-ray spectra were obtained from a circular region with a radius 
$2.5^{\prime}$ and were background subtracted using an annular region 
surrounding the SNR. The data were simultaneously fit in the 0.5 -7.5~keV band
with a non-equilibrium ionization collisional plasma model \citep{Borkowski01},
assuming a constant temperature and single ionization parameter, combined with 
ISM absorption. The best fit requires significant over-abundance of Si, S,  
 Ar and Ca with respect to the solar values of \citet{Anders89}. An additional
component is needed to model the Fe K$\alpha$ emission at 6.4~keV. 
The values obtained  are {\it k}T $\sim 1.7 $ keV, $\tau \sim 4.7 \times 10^{10}$ s cm$^{-3}$ 
and $N_{\rm H} \sim 2.6 \times 10^{22}$ cm$^{-2}$. 

To estimate the distance to G352.7-0.1, we analized HI profiles 
 in the direction to the bright 
eastern spot in G352.7-0.1. These profiles reveal an absorption HI 
feature around 
V$_{\rm LSR} \sim$ -85~km/s. A flat rotation curve of the Galaxy 
produces for this 
radial velocity a near distance of  $\sim$ 6.6~kpc and a far distance of
$\sim$ 10.2~kpc.
 Since no absorption features are observed at more negative velocities, 
6.6 kpc can be confidently set as the lower limit for the distance. 
An upper limit is the tangent point at $\sim$ -189~km/s, 
 corresponding to the  kinematical distance of 8.4~kpc. Based on  CO 
and HI observations we estimated the cumulative absorbing column density
in direction to G352.7-0.1. The obtained value is in good agreement with 
the  $N_{\rm H}$ derived
from the XMM-Newton spectrum if the SNR is located at a distance between 
6.6 and 8.4~kpc.     
On the other hand, the analysis of the distribution of the HI emission near  
  7~kpc reveals the presence of an open shell  
 surrounding G352.7-0.1. It is possible that this  
shell represents the walls of the wind-blown cavity  
 suggested by Kinugasa et al. (1998).

In summary, from the radio and X-ray study of the SNR G352.7-0.1 and its
environs it can be concluded that: 
(a) This SNR belongs to the thermal composite morphological class;
 (b) the bright radio spot seen to the
east of the SNR has not counterpart in the X-ray emission; (c) the distance
to G352.7-0.1 is set between 6.6 and 8.4~kpc; (d)  
 G352.7-0.1 is located within a cavity, probably created by 
the stellar wind of its precursor star.  

\section*{acknowledgments}

G.D. and E.G. are  grateful to the SOC of the 
``X-ray universe 2005" conference for the financial support for their attendance to the meeting.
This research was partially funded by the Argentina Grants UBACYT A055, 
ANPCyT-PICT04-14018 and PICT-CONICET05-6433.

\end{document}